\documentclass[12pt]{article}
\usepackage{amsmath}
\usepackage[T1]{fontenc}
\usepackage[utf8]{inputenc}
\usepackage{authblk}
\usepackage{slashed}

\usepackage{hyperref}
\usepackage{graphicx}
\usepackage{fancybox}
\usepackage{color}
\usepackage{epsfig,graphicx}
\usepackage{amsfonts}

\title{Schwinger pair production: Explicit Localization of the world-line instanton  }

\author[1]{James Gordon and Gordon W. Semenoff} 

\affil[1]{ Department of Physics and Astronomy, University of
British Columbia, 6224 Agricultural Road, Vancouver, British Columbia, Canada V6T 1Z1}

\date{} 
\begin{document}

\maketitle

\begin{abstract}
 We  present a simple proof that   
 the imaginary part of the world-line path integral which computes
 the  rate of  
 Schwinger pair production
of charged particles in a constant electric field,  is given exactly
by the semiclassical WKB limit.    
 \end{abstract}

  Schwinger's famous  formula \cite{Schwinger:1951nm} computes  
the probability,
$P=1-e^{-\gamma V}$,  
 of the production of charged 
particle-antiparticle pairs by a constant external electric field, 
where, for spin zero particles,    
\begin{equation}\label{schwinger}
 \gamma=\sum_{n=1}^{\infty }\frac{\left(- 1\right)^{(n+1)} E^2}{8\pi ^3n^2} e^{-\pi
 m^2n/|E|}
\end{equation}
Here $m$  is the mass of the particles, $V$ is the space-time volume   and $E$ is the electric field, into which we
have absorbed a factor of the electric charge.   
The problem of computing the damping rate  $\gamma$
can be posed as that of evaluating the imaginary part of 
the world-line path integral for the relativistic particle,
\begin{align}\label{euclidean}
\gamma = -2\Im\frac{1}{V}\int_0^\infty\frac{dT}{T }\int [dx_\mu(\tau)]e^{-\int_0^1d\tau
\left[\frac{T}{4} \dot x_\mu(\tau)\dot x_\mu(\tau)+E
 x_1(\tau)\dot x_2(\tau)\right]
-\frac{m^2}{T}}
\end{align}
The integral is over periodic paths, $x_\mu(\tau+1)=x_\mu(\tau)$ and the space-time metric has Euclidean
signature.  Although the imaginary part of this integral can be readily be found to be (\ref{schwinger})
by integrating the Gaussian 
variables $x_\mu(\tau)$ and then finding the imaginary part of the remaining integral over $T$,  
it is often convenient to perform a first principles semi-classical evaluation
   of the integral in (\ref{euclidean}).  In that case, one treats $x_\mu(\tau)$ and $T$ as dynamical variables,
   solves the classical equations of motion, expands in fluctuations about the solution and integrates over the 
   fluctuations.  In fact, this second approach is needed for more general cases where the gauge field is not just a constant
   electric field \cite{Dunne:2006ff} or when the gauge field is dynamical \cite{Semenoff:2011ng}.  Then the action is not just quadratic in the
   coordinates but has higher order terms and the saddle point approximation to the world-line path integral is needed.  
   In the following, shall have nothing to say about the more general case.  What we demonstrate is that the
   the semiclassical approximation to  (\ref{euclidean}) is not an approximation.  It gives the exact result. 
   
   An electric field in 
   Euclidean space resembles a magnetic field and the classical instanton solutions are cyclotron orbits in that field \cite{Affleck:1981bma}.   The fluctuations
   about those solutions have tachyonic modes which make the instanton amplitudes imaginary and give (\ref{euclidean}),
   which is nominally an integral of a real integrand over real variables, an imaginary part.   (This integral gets its imaginary
   part from the fact that the quadratic form in $x_\mu$ is not positive for all values of $T$ and, for values of $T$ 
   where it has negative eigenvalues, it has to be
   defined by analytic continuation.)

The semiclassical computation treats both $x_\mu(\tau)$ and $T$ as dynamical variables which, to leading order, solve the classical equations of motion derived from the world-line action, 
\begin{align}
S=\int_0^1 d\tau \left[ \frac{T}{4}\dot x_\mu(\tau)^2 +Ex_1(\tau)\dot x_2(\tau)+\frac{m^2}{T}\right]
\label{world-lineaction}
\end{align}
The classical equations of motion are
\begin{align}
\frac{1}{4m^2}\int_0^1 \dot x^2&=\frac{1}{T^2}
~,~-\frac{T}{2}\ddot x_1 -E\dot x_2=0 ~,~
-\frac{T}{2}\ddot x_2  +E\dot x_1=0~,~
-\frac{T}{2}\ddot x_{3,4}&=0
\end{align}
with periodic boundary conditions, $x_\mu(\tau+1)=x_\mu(\tau)$. 
The solutions of these equations are
\begin{align}\label{classical_solution}
x^{(n)}_{0\mu}(\tau)=\frac{m}{E}(\cos2\pi n\tau,\sin2\pi n\tau,0,0)
  ~,~
T_0^{(n)}=\frac{E}{\pi n}~,~n=1,2,3,...
\end{align}
where $n$ is the instanton number.  In the $n$-instanton sector, the integration variables in (\ref{euclidean}) are the classical solutions plus
fluctuations, 
\begin{align}
x_\mu(\tau)=x^{(n)}_{0\mu}(\tau)+\delta x_\mu(\tau)
~,~T=T_0^{(n)}+ \delta T
\end{align}
and we shall expand the action  in the fluctuations. 

Before we proceed, we note that, in order to make the quadratic truncation of this theory well-defined, we need to deal with a collective coordinate.   If $x^{(n)}_{0\mu}(\tau)$ is a solution of the classical equation of motion, so is $x^{(n)}_{0\mu}(\tau+t)$ and this degeneracy of solutions leads to a zero mode of the linearized equations of motion for the fluctuations. To fix this, we introduce a collective coordinate by inserting the identity
\begin{align}
1=\frac{1}{\omega} \int_0^1 dt  \delta\left( \int_0^1 d\tau\left[ \dot x^{(n)}_{0\mu}(\tau)  x_\mu(\tau+t)\right]\right)
\left| \frac{d}{dt} \int_0^1 d\tau  \dot x_\mu (\tau)\delta x_\mu(\tau+t)\right|
\end{align}
into the path integral.  In this expression, $\omega=2n$ is the number of Gribov copies, that is, the number of solutions for the variable $t$ of the equation \mbox{$ \int_0^1 \!d\tau\!\left[ \dot x^{(n)}_{0\mu}(\tau-t) x_\mu(\tau)\right]=0$} in the interval $t\in[0,1)$. 
Upon using translation invariance of the integrand and measure, we can see that this is equivalent to inserting the gauge fixing factor
\begin{align}\label{gf}
\int[dbdcd\bar c]~e^{\int_0^1d\tau\left[ 2\pi i\omega b\dot x^{(n)}_{0\mu}(\tau) \delta x_\mu(\tau)
+  \bar c \dot x^{(n)}_{0\mu}(\tau)\dot  x_\mu(\tau)c\right]}
\end{align}
into the path integral.  Here, the $\tau$-independent variables $c$ and  $\bar c$  are anti-commuting Faddeev-Popov ghosts and $b$ is a  Lagrange multiplier. 

We add the exponent in the integrand of equation (\ref{gf}) to the classical action $S$ from equation (\ref{world-lineaction}) to get the ``gauge fixed action '' $S_{\rm gf}$. The world-line path integral is now
\begin{align}\label{euclidean1}
	i\gamma = -2\sum_{n=1}^\infty\frac{1}{V}
~\int ~ [d(\delta x_\mu(\tau))d(\delta T)dbdcd\bar c ]~
\frac{1}{T_0^{(n)}+\delta T}~e^{-S_{\rm gf}[x^{(n)}_{0\mu}+\delta x_\mu,T_0^{(n)}+\delta T,b,c,\bar c] 
}
\end{align}
where, because it turns out that each instanton sector has an odd number of tachyons, the Euclidean partition function is purely imaginary in all of the instanton sectors (so we have removed the symbol $\Im$).
When it is expanded about the $n$-instanton solution, the action contains classical, quadratic and higher order (interaction) terms,
\begin{align}
S_{\rm gf}=S^{(n)}_{\rm classical}+S^{(n)}_{\rm quad}+S^{(n)}_{\rm int}
\end{align}
respectively, and 
\begin{align}
&S^{(n)}_{\rm classical}=\int_0^1 \!d\tau \left[ \frac{T_0^{(n)}}{4}\dot x^{(n)}_{0\mu}(\tau)^2 +Ex^{(n)}_{01}(\tau)\dot x^{(n)}_{02}(\tau)+\frac{m^2}{T_0^{(n)}}\right] = \frac{\pi n m^2}{E}  \label{classical_action} \\
&S^{(n)}_{\rm quad}= \int_0^1 \! d\tau \left[ \frac{\delta T}{2}\dot x^{(n)}_{0\mu} \delta\dot x_\mu+\frac{T_0^{(n)}}{4}\delta\dot x_\mu^2
 +E\delta x_1 \delta\dot x_2 +\frac{m^2\delta T^2}{T_0^{(n)3}}
 -2\pi i\omega b\dot x^{(n)}_{0\mu}  \delta x_\mu -  \bar cc \dot x^{(n)}_{0\mu} \dot  x^{(n)}_{0\mu} \right]
 \label{0_action}  \\
&S^{(n)}_{\rm int}= 
\int_0^1 \!d\tau\left[  \frac{1}{4} \delta T\,\delta\dot x_\mu(\tau) \delta\dot x_\mu(\tau) - \dot x^{(n)}_{0\mu}(\tau)~ \bar cc \delta \dot  x_\mu(\tau)\right]
+\sum_{k=3}^\infty  \frac{  m^2}{T_0^{(n)}} \frac{(-\delta T)^k}
{T_0^{(n)k}}  
\label{int_action}
\end{align}
It is in the remaining integral that we want to show that the interaction terms, $S_{\rm int}$,  and the term in the measure, $1/(T_0^{(n)}+\delta T)$, can be replaced by zero and $1/T_0^{(n)}$, respectively. 
 
For this purpose, we define the fermionic transformation 
\begin{align}\label{brstsusy}
 \Delta\bar c=\frac{1}{2} \delta T ~,~
 \Delta ( \delta x_\mu (\tau))= c x^{(n)}_{0\mu}(\tau)~,~\Delta c=0~,~\Delta b=0~,~\Delta(\delta T)=0
\end{align}
We note that 
\begin{align}
\Delta^2({\rm anything})=0 ~~,~~\int  [d(\delta x_\mu(\tau))d(\delta T)dbdc d\bar c ]~\Delta({\rm anything})=0
\end{align}
and the three parts of the action are invariant separately, 
\begin{align}
 \Delta S^{(n)}_{\rm classical}=0~,~\Delta S^{(n)}_{\rm quad} =0~,\Delta S^{(n)}_{\rm int}=0
  \end{align}
Moreover, the interaction terms in the action and in the measure can be seen to be exact, that is, they are   
\begin{align}
 &S^{(n)}_{\rm int}=\Delta\psi   \\
&\psi= \frac{\bar c}{2} \int_0^1d\tau   \delta\dot x_\mu(\tau)^2 -2\bar c\sum_{k=3}^\infty 
   \frac{ m^2}{T_0^{(n)}}  \frac{(-\delta T)^{k-1}}{T_0^{(n)k}} 
   \\
 &  \frac{1}{T_0^{(n)}+\beta \delta T}= \frac{1}{T_0^{(n)} }+\Delta \chi \\
&   \chi = \bar c   \sum_{k=1}^\infty \frac{ (-\beta)^k \delta T^{k-1} }{ T_0^{(n)k} }
\end{align}
Now, consider 
$$
{\mathcal I}(\beta,\lambda)\equiv \int [d(\delta x_\mu)d(\delta T)dbdcd\bar c]
\left(  \frac{1}{T_0^{(n)}+\beta\delta T}\right)~ e^{-S^{(n)}_{\rm classical}-S^{(n)}_{\rm quad}-\lambda S^{(n)}_{\rm int}}
$$
We are interested in this integral when the parameters $\beta=1$ and $\lambda=1$.  However, we can easily
show that it is independent of both $\beta$ and $\lambda$.   Consider
$$
\frac{\partial}{\partial\lambda}{\mathcal I}(\beta,\lambda)=\frac{d}{d\lambda}\int [d(\delta x_\mu)d(\delta T)dbdcd\bar c]
\left(  \frac{1}{T_0^{(n)}}+  \Delta \chi\right)~ e^{- S^{(n)}_{\rm classical}-S^{(n)}_{\rm quad}-\lambda \Delta \psi}
$$
$$
=- \int [d(\delta x_\mu)d(\delta T)dbdcd\bar c]  \left\{
\left(  \frac{1}{T_0^{(n)}}+  \Delta \chi\right)  \right\}\Delta\psi ~e^{- S^{(n)}_{\rm classical}-S^{(n)}_{\rm quad}-\lambda \Delta( \psi)}
$$
$$
=- \int [d(\delta x_\mu)d(\delta T)dbdcd\bar c ]~ \Delta\left\{
\left(  \frac{1}{T_0^{(n)}}+  \Delta \chi \right) \psi e^{- S^{(n)}_{\rm classical}-S^{(n)}_{\rm quad}-\lambda\Delta \psi}  \right\}=0
 $$
 Similarly, 
 $$
 \frac{\partial}{\partial\beta}{\mathcal I}(\beta,\lambda)
= \frac{\partial}{\partial\beta}\int [d(\delta x_\mu)d(\delta T)dbdcd\bar c]
\left(  \frac{1}{T_0^{(n)}}+ \Delta \chi\right)~ e^{-S^{(n)}_{\rm classical}-S^{(n)}_{\rm quad}-\lambda S^{(n)}_{\rm int}}
$$
$$
= \int [d(\delta x_\mu)d(\delta T)dbdcd\bar c]
~ \Delta \left( \frac{d}{d\beta}\chi\right)~ e^{-S^{(n)}_{\rm classical}-S^{(n)}_{\rm quad}-\lambda S^{(n)}_{\rm int}}
$$
$$
= \int [d(\delta x_\mu)d(\delta T)dbdcd\bar c]
~ \Delta\left\{\left( \frac{d}{d\beta}\chi\right)~ e^{-S^{(n)}_{\rm classical}-S^{(n)}_{\rm quad}-\lambda S^{(n)}_{\rm int}}\right\}=0
$$ 
and the integral 
$$
{\mathcal I}(\beta,\lambda)=  \int [d(\delta x_\mu)d(\delta T)dbdcd\bar c]\left(  \frac{1}{T_0^{(n)}+\beta\delta T}\right) 
e^{-S^{(n)}_{\rm classical}-S^{(n)}_{\rm quad}-\lambda S^{(n)}_{\rm int}}
$$
is independent of the parameters $\lambda$ and $\beta$.  Both of these parameters can then be deformed to zero, yielding
\begin{align}\label{euclidean2}
i\gamma = -2\sum_{n=1}^\infty
 \frac{1}{V}~\int ~ [d(\delta x_\mu(\tau))d(\delta T)dbdcd\bar c ]~
\frac{1}{T_0^{(n)} }~e^{- S^{(n)}_{\rm classical}-S^{(n)}_{\rm quad}
}
\end{align} 
which, in each instanton sector, contains the classical solution plus quadratic fluctuations only.  The remaining integral was performed in reference \cite{Gordon:2014aba}
where zeta-function regularization was used to define the infinite sums and products which appear.  The result
yields the known Schwinger formula (\ref{schwinger}).  

In a similar vein, it is possible to show that the imaginary part of the sigma model functional integral for oriented open strings in a constant electric field is given exactly by the WKB limit of the semi-classical expansion about world-sheet instantons, and  summing over  
instanton number \cite{Gordon:2016}.

\vskip 0.5cm

\noindent
The authors acknowledge financial support of NSERC of Canada.  We thank James P. Edwards for comments
on the manuscript.

\end{document}